# A systematic approach to answering the easy problems of consciousness based on an executable cognitive system

Qi Zhang

San Jose, California, USA

## Abstract

Consciousness is the window of the brain and reflects many fundamental cognitive properties involving both computational and cognitive mechanisms. A collection of these properties was described as the "easy problems" by Chalmers, including the ability to discriminate, categorize, and react to stimuli; information integration; reportability; information access; attention; deliberate control; and the difference between wakefulness and sleep. These "easy problems" have not been systematically addressed. This study presents a first attempt to address them systematically based on an executable cognitive system and its implemented computational mechanisms, built upon an understanding of conceptual knowledge proposed by Kant. The study suggests that the abilities to discriminate, categorize, react, report, and integrate information can all be derived from the system's learning mechanism; attention and deliberate control are goal-oriented and can be attributed to emotional states and its information-manipulation mechanism; and the difference between wakefulness and dream sleep lies mainly in the source of stimuli. The connections between the implemented mechanisms in the executive system and conclusions drawn from empirical findings are also discussed, and many of these discussions and conclusions are supported by demonstrations of the executive system.

# Introduction

Consciousness, in its simplest terms, refers to one's subjective experience, and its nature remains largely unknown. However, even under this simple definition, consciousness encompasses a broad range of mental phenomena. The wonder of consciousness has been a subject of inquiry into the soul for thousands of years, only transitioning into scientific investigation in the middle of the last century. The most notable driving forces behind this transition include the quest for the neural correlates of consciousness (NCC) initiated by Crick and Koch (1990), technological advancements in neuroscience—such as fMRI and EEG—and Chalmers' (1995) introduction of the distinction between the easy and hard problems of consciousness, which has reignited a fresh wave of scientific and philosophical debates.

The easy problems of consciousness refer to phenomena that can be "explained in terms of computational or neural mechanisms," and Chalmers listed seven of them:

- The ability to discriminate, categorize, and react to environmental stimuli
- The integration of information by a cognitive system
- The reportability of mental states
- The ability of a system to access its own internal states
- The focus of attention
- The deliberate control of behavior
- The difference between wakefulness and sleep

Thirty years later, the easy problems have hardly been answered directly and systematically. This slow progress aligns with Chalmers' argument that we "do not yet have anything close to a complete explanation of these phenomena" because "they are among the most interesting unsolved problems in cognitive science" and "getting the details right will probably take a century or two of difficult empirical work."

A great number of theories of consciousness have been proposed, of which about twenty are considered either neurobiological or potentially expressible in neurobiological terms, as reviewed by Seth and Bayne (2022). Among these, a few are generally regarded as the most influential and have appeared in many reviews and studies. These include different versions of global workspace theories (GWTs) (Baars 1988, Dehaene and Changeux 2011, Mashour et al. 2020), recurrent processing theory (RPT) (Lamme and Roelfsema 2000, Lamme 2006), different versions of higher-order thought theories (HOTs) (Rosenthal 2005, Lau and Rosenthal 2011, Brown et al. 2019), and integrated information theory (IIT) (Tononi 2008, Tononi et al. 2016). These theories are generally conceptual and lack sufficient computational or neural mechanisms to tackle the easy problems directly and systematically, although some provide more functional and cognitive details than others. Furthermore, each theory has its unique core claims, as phrased by Seth and Bayne (2022), regarding the factors that give rise to consciousness, which often distinguish one theory from another. A unified theory is still sought to uncover the nature of consciousness as a whole and, therefore, to systematically address both the easy and hard problems together.

Revealing the mechanisms, or the functional and cognitive details needed to answer the easy and hard problems, is a challenging task, especially when they are expected to align with neurobiologically based findings from empirical studies. The most notable example is the difficulty researchers have faced in searching for the NCC. An NCC is defined as the minimal set of neural events jointly sufficient for a conscious state. It is thought to involve the activation of neural regions in the brain relevant to that state, which can be identified through carefully designed and conducted empirical studies. However, these efforts have been severely challenged by the temporal dynamics of cortical activations, making it difficult to distinguish the NCC proper from the neural prerequisites and consequences of consciousness (Aru et al. 2012; Tsuchiya et al. 2015; Koch et al. 2016). This challenge has even led to the proposal of multiple NCCs (Block 2005).

On the other hand, a wide range of insights and mechanisms have been proposed and derived from both theoretical and empirical studies across various fields and disciplines within cognitive and neuroscience research. These studies often focus on specific cognitive functions and mechanisms. We can find a substantial collection of findings that are directly relevant to one or more of the easy problems, with some addressing a single problem and others applicable to multiple problems. For example, the abilities to discriminate and categorize may result from a single mechanism (Goldstone 1994) or distinct mechanisms (Peelen and Downing 2007); information integration may occur through hierarchical processing (Felleman and Van Essen 1991) or synchronized feature binding (Singer 1999); and deliberate control may be exerted via the prefrontal cortex (Baddeley 1996) or a feedback loop (Carver and Scheier 1998).

It is clear that a systematic understanding of the biological brain is still lacking in answering all the easy problems, given the many unknowns that exist between the well-defined neuron model (Hodgkin and Huxley 1952) and the well-studied high-level cognitive capacities and phenomena of the biological brain, particularly the human brain. However, an alternative approach may be to employ an artificial cognitive system to explain all the easy problems within a single framework—provided that the system possesses sufficient mechanisms to address some or most of these problems. This approach is valid if the employed system meets the definition of a cognitive system, meaning it processes information in ways similar to human cognition when performing cognitive tasks such as learning, responding, reasoning, and problem-solving (Vernon 2021).

This study attempts to systematically address most of the easy problems based on a cognitive system, named adder, developed by the author. The adder is an executable cognitive system that can be employed to demonstrate and support discussions in answering the easy problems. The system is a continuous development of the author's previous work on cognitive systems across several studies: the mechanism for learning from experience by grounding symbols in real-world meanings (Zhang 2005) to solve the symbol grounding problem (Harnad 1990); the mechanism for acquiring semantic knowledge during dream sleep (Zhang 2009a), which is characterized by the random firing of past experiences (Hobson et al. 2000); and the cognitive construction of semantic and episodic memories in a cognitive system (Zhang 2011).

In this study, the cognitive construct of the adder is introduced, with a focus on its declarative memory, which consists of both episodic and semantic memory, as well as its mental operator (its equivalent of the prefrontal cortex, PFC). The easy problems are addressed in their original order by: (1) briefly reviewing several representative theories and empirical findings that share the same scope of interest, including influential theories of consciousness; (2) elaborating on the mechanisms implemented in the system, explaining their relevance to the easy problems, and demonstrating these mechanisms through semantic and image-based reporting in response to external stimuli; and (3) aligning the mechanisms offered by the adder with theories and findings from referenced studies.

## The adder, its cognitive construct and semantic memory

First, we introduce the overall cognitive structure of the adder, including its multiple cognitive regions and how they are connected through information flows to enable cognitive capacity at the system level. Then, we focus on its semantic memory regions and the mechanisms of semantic knowledge acquisition and application. This work aims to introduce the overall system, its cognitive regions, and related mechanisms with minimal computational details, although many of these details can be found in the author's previous studies.

**The cognitive construct**

The adder can learn a small amount of conceptual knowledge from its experiences and subsequently respond to external stimuli using that knowledge. It can learn to recognize simple objects in images, such as lines, squares, and rectangles, and perform basic operations like counting objects or adding and subtracting them. The term "simple" is used from a human perspective, as five-year-old children have typically already mastered these mental operations with numbers up to ten (Andres and Pesenti 2014).

However, these tasks remain impossible for even the most advanced generative AI systems today, such as ChatGPT, Copilot, and Gemini. For example, if one uploads an image containing eight or nine objects to any of these AI systems and asks them to count the objects, the response (1) is incorrect at an unacceptably high rate and (2) lacks consistency. OpenAI explicitly warns users not to trust ChatGPT's counting ability, stating on its homepage that ChatGPT "may give approximate counts for objects in images" (OpenAI 2025).

The adder can learn simple concepts from experience, respond to stimuli, and perform mental operations when prompted. These capacities are rooted in mechanisms and cognitive processes that are central to the easy problems, such as discrimination, categorization, reaction to environmental stimuli, information integration, and reportability. Fig. 1 illustrates the cognitive construct abstracted from the adder, which consists of several cognitive regions responsible for declarative memory, emotion, and mental operation. Its declarative memory comprises its semantic memory (a combination of the symbol subsystem (SS) and feature subsystem (FS) in Fig. 1) and its episodic memory. The former is the region that learns conceptual knowledge and responds to stimuli based on acquired knowledge, while the latter stores past experiences for later processing. The operator functions as the adder's equivalent of the PFC for mental operations to (1) extract operational knowledge and (2) manipulate information based on acquired operational knowledge. The boundary of each cognitive region serves as its interface for information transformation and redistribution between regions. All cognitive regions in Fig. 1 are functional, except for the emotion block, which is nonfunctional and included only to acknowledge the importance of emotion as an internal driving force.

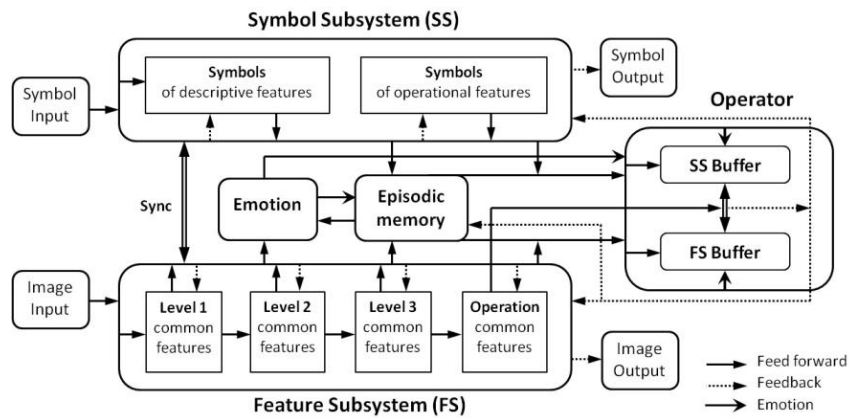

Figure 1. The cognitive construct and information flow of the adder. It has cognitive and functional regions for its semantic memory (the SS and FS pair), episodic memory, and mental operator. It receives external stimuli through one or both input portals and gives off its semantic or image report output at the output portals.

**Human declarative memory**

Human memory is the cognitive ability to encode, store, and retrieve information, serving as a fundamental function that underpins learning, decision-making, and problem-solving. The answers to the easy problems are closely tied to the acquisition, categorization, storage, and retrieval of memory, particularly declarative memory or declarative knowledge. Some essential properties of declarative memory are summarized as follows. Human long-term memory is divided into declarative and nondeclarative memory (Graf and Schacter 1985). Nondeclarative memory encompasses priming, perceptual learning, and procedural skills, while declarative memory consists of semantic and episodic memory (Squire 2004, Tulving 1983). Episodic memory pertains to events, characterized by their temporal sequence and spatial context, whereas semantic memory involves factual and conceptual knowledge that is independent of specific past experiences.

Newly acquired episodic memory is generally associated with the medial temporal lobe, including the hippocampus and its surrounding cortices, whereas semantic memory resides in the general neocortex and is independent of the medial temporal lobe (Eichenbaum 2004). The development of semantic memory depends on the retention of episodic memory and is most likely derived from episodic memory through a cognitive process called memory consolidation. The relationship between the acquisition of semantic and episodic memory is twofold. First, episodic memory may be a prerequisite for the formation of semantic memory. Amnesic patients who have lost the ability to retain episodic memory often exhibit impaired semantic memory; for example, they struggle to acquire new semantic knowledge (Squire and Zola 1998). However, amnesic patients may still retain intact implicit memory and previously acquired semantic knowledge (Cohen and Squire 1980). Additionally, impairment of episodic memory in early childhood often results in severe mental retardation, which is believed to stem from the obstruction of semantic learning in the absence of episodic memory (Baddeley et al. 2001). Second, the factual knowledge is most likely abstracted and consolidated from past experiences or various episodic memories.

**The episodic memory of the adder**

The episodic memory of the adder corresponds to the episodic memory block in Fig. 1, which consists of multiple parallel memory units, each designated to store a segment of an external experience (see Zhang 2009a, 2009b). These units are interlocked, ensuring that the sequential segments of an external experience are stored in their original order of occurrence. The interlocking mechanism functions similarly to the hippocampus in terms of sequential learning when the variable is time (Levy 1996) and spatial navigation when the variable is distance (Muller et al. 1987). As a result, all stored segments of experiences can be re-accessed and replayed either in their original sequence or randomly, independent of their original order.

In the author's previous studies, it has been demonstrated that semantic memory can be acquired through either sequential firing or random firing, although the learning rate in sequential firing is significantly higher than in random firing. Since (1) random firing of episodic memory is generally considered a cause of dreaming (Hobson 1988, Wolf 1994), and (2) the replaying of recent waking patterns of neuronal activity within the hippocampus has been observed during dream sleep (Louie and Wilson 2001), this previous study is regarded as a computational demonstration of the role of dream sleep in learning and memory consolidation, as concluded in many empirical studies (Pearlman 1971, Bloch et al. 1979). We will revisit this learning mechanism when addressing the seventh subject of the easy problems.

**The semantic memory of the adder**

Semantic memory refers to an individual's knowledge of facts and concepts, which is acquired through experience. As a component of declarative memory, semantic knowledge can always be expressed or communicated in a language that can be commonly shared with others. In other words, semantic knowledge represents the meanings of things as understood by an individual from specific experiences, and these meanings can be shared with others through common language. While humans have the cognitive capacity to develop the most sophisticated semantic knowledge on Earth, studies indicate that many mammals, especially primates, as well as certain bird species, are also capable of acquiring semantic knowledge and using their language to communicate with one another within their populations (Martín-Ordás, 2021).

The mechanism of semantic memory in the adder and its predecessors is built upon the understanding proposed by I. Kant (1787/1998) that a concept is a common feature or characteristic, and concepts are abstracts in that they omit the differences of the things in their extension, treating them as if they were identical.  He further explained that a common feature is "a mark, which can be common to many things". In other words, one must focus on the relevant features in the experience and ignore those that are irrelevant (Bourne et al. 1986). Therefore, semantic knowledge is equivalent to and can be expressed with a set of common features or characteristics, and knowledge acquisition becomes a process of sensing, extracting, and storing these common features.

The semantic memory of the adder is realized through paired cognitive regions: the feature subsystem (FS) and the symbol subsystem (SS), as shown in Fig. 1. This paired construct has also been a standard framework throughout the author's previous studies. The FS implements Kant's principle of common features, enabling it to extract, learn, and retain the shared characteristics (meanings) of things. On the other hand, the SS is responsible for learning and retaining the symbolic representations of the common features stored in the FS. The pairing of symbols and common features is synchronized during learning, indicated as "Sync" in Fig.1, following the principle that "what fires together, wires together" (Hebb 1949). As a result, information stored in both the FS and SS shares the same internal code or state. Therefore, activating information in one subsystem can trigger the corresponding information in the other subsystem.

Semantic knowledge can be further divided into two main components: factual knowledge and operational knowledge. While factual knowledge refers to the "what," operational knowledge pertains to the "how" of understanding and reasoning—i.e., what things are (facts or descriptions) and how things work (processes). Simple facts are descriptive in nature, whereas operational knowledge involves actions characterized by procedures and causal relationships. Similarly, the semantic memory in the adder contains both common features of objects (descriptive features) and operations (operational features). Its descriptive features are organized in a multi-level hierarchy, with common features connected in parallel within one level. In Fig. 1, level 1 represents the lowest level of concepts, learning the common features conveyed by sensory input, while level 2 learns the common features derived from those in level 1. On the other hand, its operational features are extracted by the operator and then retained in the FS.

**The operator**

The operator, as shown in Fig. 1, is the adder's equivalent of the PFC. It is the cognitive region that perceives and extracts operational rules connecting cause to effect, as well as manipulates information based on acquired operational knowledge before applying the operation or its result to the external world. It also serves as a hub where information from the cognitive regions of the SS, FS, episodic memory, and emotion converges and interacts. For example, the operator may take inputs activated from the FS of the semantic memory and manipulate them based on operational knowledge that is also activated from FS, i.e., operational common features. It then projects the result of this manipulation back to both the SS and FS, triggering symbol or image output. Some of these operations will be demonstrated in later discussions.

The operator consists of a synchronized pair of buffers that can extract actions or operations from a sequence of information using computational mechanisms to compare and identify differences that always exist (i.e., operational common features) in operations associated to same symbol. It may then manipulate given information based on the operational common feature once activated by external input. In other words, the operator has two main functions. The first is to extract common features of operations from sequences of information with temporal associations and causal relationships. The second is to apply the acquired common features of operations to manipulate information within the operator. These functions are considered to align with similar functions attributed to the PFC of humans and other highly intelligent primates, the less developed PFC of other mammals, the nidopallium caudolaterale (NCL) of bird brains, and the dorsal ventricular ridge (DVR) of reptiles (Eugen et al. 2020, Preuss and Wise 2022).

**Emotion and attention**

Emotions are both physical and mental states that arise in the limbic system and are believed to have emerged with the rise of reptiles, long before mammals. Emotions have evolved to serve various adaptive functions and play an important role in survival and well-being, from threat detection, decision-making, and motivation to memory enhancement and social communication (LeDoux 1996, Buss 2015). In D. Hume's view, our actions are motivated by "fears, desires, and passions," and even reasoning serves emotions—highlighting the importance of emotions to highly intelligent organisms.

Emotion and attention are closely linked, with emotions significantly influencing where and how we focus our attention. Emotionally salient stimuli tend to automatically capture our attention more readily than neutral stimuli (Brosch et al. 2013). Moreover, emotions can either narrow our focus on a specific threat or broaden our attention to enhance the survival and well-being. In addition, the importance of attention has been well recognized in consciousness studies, and it has been proposed to serve a gating function in some theories of consciousness (e.g., Baars 1988, Dehaene and Changeux 2011).

Unlike semantic memory and the operator, the emotion block in Fig. 1 is not implemented with a functional mechanism. This block and its connections to the two buffers in the operator indicate recognition of the importance of emotion in influencing and biasing the system's learning and mental operations. Similarly, its connection to episodic memory acknowledges that emotions can be attached to episodic memory, becoming part of the information manipulated in the operator. Furthermore, although this study agrees with the proposed function of attention as a filter or gate, it does not designate a specific region for attention or implement a mechanism for it. However, since the boundary of each functional region in the construct derived from the interface for information transformation and redistribution between regions, the operator's interface can incorporate a filtering function.

# Answering the easy problems

While addressing each of the easy problems, we begin by briefly reviewing representative theories and empirical findings from both influential theories of consciousness and other cognitive and neuroscience studies that share the same scope of interest as the specific easy problem. We then answer and explain the easy problem primarily through the mechanisms implemented in the adder, often demonstrating them via the executable system's response to external stimuli. Finally, we highlight the theories and findings that closely align with the mechanisms proposed by the adder.

In this study, the adder, equipped with previously learned knowledge, is employed to demonstrate its functions and mechanisms by responding to external stimuli in relation to the discussed easy problem. The adder is capable of learning a few concepts of both descriptive and operational knowledge from various external stimuli experienced earlier. The descriptive knowledge includes simple patterns or objects presented in image form, while the operational knowledge encompasses actions of information manipulation such as counting, adding, or subtracting, which occur in the mental operator. Details about the system's learning processes can be found in the author's previous studies.

**The ability to discriminate, categorize, and react to environmental stimuli**

The abilities to discriminate, categorize, and react are fundamental to how the brain processes and makes sense of the world, learns new concepts, and responds effectively to information. Whether these abilities arise from a single underlying mechanism or distinct mechanisms remains an ongoing topic of research. The unified mechanism view proposes that these abilities stem from a shared neural and computational mechanism in the brain (Goldstone 1994), whereas the distinct mechanism view suggests that separate neural circuits underlie discrimination, categorization, and reaction processes (Peelen and Downing 2007).

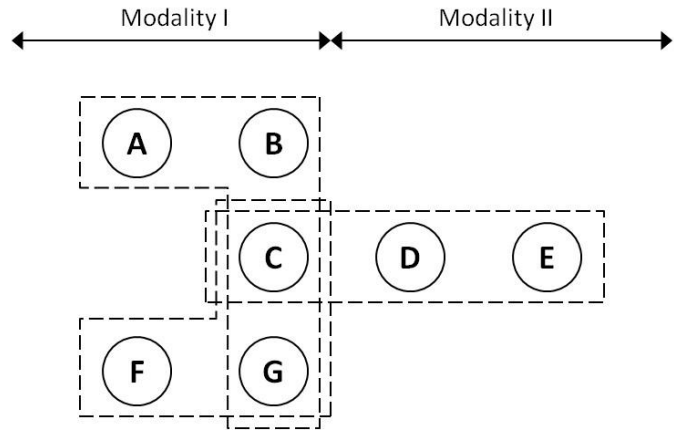

Figure 2. Common features and their role in discrimination, categorization, and integration. Three concepts (i.e., mental representations of categories) are indicated by multiple common features: ABCG, CDE, and CGF. When an external input best matches a concept (category) and activates it, this activation inhibits (discriminates) other competing activations. Since common features are rooted in their origins (e.g., modality-specific origins), the activation of a concept is the consequence of integrated activations of its subordinate factors or common features.

Theories of consciousness generally do not address the cognitive mechanisms involved in discriminating, categorizing, and reacting to external stimuli, except to consider them low-level processes supported by empirical findings. On the other hand, beyond confirming their low-level nature, empirical studies have provided additional insights. For example, stimuli are discriminated based on modality related features such as color, shape, or sound frequency, and they are categorized based on shared characteristics or learned associations, facilitating efficient processing and decision-making (Fedota et al. 2012, Bryan et al. 2024).

The mechanism underlying the ability to discriminate, categorize, and react in the adder aligns with the unified perspective of a shared mechanism. A shared mechanism not only easily explains why these abilities are closely intertwined but is also the most energy- and computation-efficient, which is plausible in terms of the stable equilibrium defined by the principle of minimum energy in physics, a principle that applies to all stable systems in nature. Similar to empirical findings and the general understanding of the influential theories of consciousness, the adder's abilities also operate at "lower" levels within both sensory and semantic knowledge regions, below the highest level in the mental operator.

Table 1. Examples of the adder perceiving and counting objects in image

| Image input | Common features | | Symbol output |
|---|---|---|---|
| | Descriptive | Operational | |
| 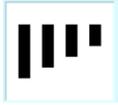 | V: vertical line | Count | 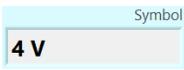 4 V |
| 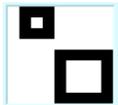 | S: 4 equal sides, adjacent sides perpendicular | Count | 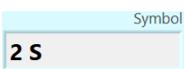 2 S |
| 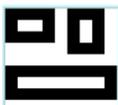 | R: 4 sides, opposite sides equal, adjacent sides perpendicular | Count | 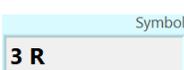 3 R |

In the adder, the shared mechanism is not an independent, stand-alone system. Rather, it is a consequence of its learning mechanism, which relies solely on common features as the criterion for learning and response. This shared mechanism is rooted in the acquired semantic knowledge and the knowledge structures built from it—i.e., the already acquired common features inherently enable discrimination, categorization, and reaction without additional support. The learning mechanism implements Kant's (1787/1998) understanding of concept knowledge, which defines a concept as a common feature or characteristic. Concepts are abstract in that they omit the differences among the things they encompass, treating them as if they were identical. Based on the principle of common features (see also Fig.2): (1) categorization occurs when entities share the same set of common features; (2) discrimination arises when one concept does not share all its common features with another; and (3) responding to external stimuli involves seeking common features in stimuli for learning or matching against known semantic knowledge. Thus, discrimination becomes a prerequisite for categorization, while categorization enhances discrimination. These abilities emerge as inherent properties of acquired semantic knowledge, which in turn reinforce continuous learning.

Table 1 summarizes three examples of how the adder utilizes common features to comprehend image inputs of external stimuli. The first column in Table 1 represents the external stimuli or images containing multiple objects. The last column shows it's symbolic outputs, which serve as a "semantic report" of what the adder has comprehended from the stimuli. The middle columns name the descriptive and operational common features. For example, in the second example, the adder perceives that there are two objects in the image based on its operational knowledge of counting. It also identifies both objects as squares due to their common features: four equal sides and adjacent sides intersecting perpendicularly. The best match between the object in the image and the concept of "S" inhibits the activation of other possibilities with less ideal matches.

Table 2. Examples of the adder responding to symbols and drawing asked objects

| Symbol input | Common features | | Image output |
|---|---|---|---|
| | Operational | Descriptive | |
| Symbol<br>5 H | Count to 5 | Horizontal line | 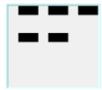 |
| Symbol<br>2 R | Count to 2 | Rectangle | 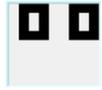 |
| Symbol<br>3 S | Count to 3 | Square | 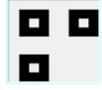 |

The ability to discriminate and categorize is expressed not only in perceiving stimuli that carry common features but also in applying acquired common features to understand symbols. Table 2 summarizes three examples of how the adder utilizes acquired common features to express its understanding of given symbols. For example, in the first example, "H" activates its associated common features, which distinguish it from other sets of common features, such as those of "V" or "S." Similarly, "5" activates the concept of counting to a total of five. As a result, five horizontal lines are drawn rather than a different number or type of objects.

**The integration of information by a cognitive system**

There are multiple views on how information may be integrated in the brain, with the most notable being hierarchical processing and neural synchronization. The former suggests that information integration occurs through multiple cortical areas and successive layers of feature detectors in a hierarchical manner (Felleman and Van Essen 1991, Riesenhuber and Poggio 1999), while the latter proposes that integration occurs through synchronized feature binding (Singer 1999, Engel et al. 2001). Furthermore, recent neuroimaging studies suggest that integration may take place in the PFC, for example, by combining visual and verbal information (Ferrari et al. 2016). On the other hand, although most influential theories of consciousness primarily address high-order integration that may give rise to consciousness, some, such as certain versions of the GWT and PRT, emphasize the importance of the PFC in integration.

Table 3. Examples of the adder to add and subtract

| Symbol input | Common features | | Symbol output |
|---|---|---|---|
| | Descriptive | Operational | Last |
| Symbol<br>4 + 3 = | Numeric | Add, count | Symbol<br>7 |
| Symbol<br>6 - 3 = | Numeric | Subtract, count | Symbol<br>3 |
| Symbol<br>9 - 2 = | Numeric | Subtract, count | Symbol<br>7 |

The views of hierarchical processing, feature binding mechanisms, and high-order integration are more complementary than conflicting. It is plausible to combine them into a comprehensive understanding that, in the biological brain, information integration occurs at multiple cognitive levels, from primary sensory cortices to associative cortices, all the way up to high-order integration. In other words, information integration is likely to occur at every cognitive step, starting from the lowest level of sensory integration, progressing through perception, and culminating in the PFC integration at the highest level—just as seen in the adder system.

The adder integrates information at all cognitive levels, from sensory processing to top-level processing in the operator. At lower levels, information integration is always associated with the integration of common features, which may be distributed across different cognitive regions, such as the regions of multiple modalities (see also Fig. 2). At the top level, however, integration often involves temporally sequenced information and may occur only in the operator. For example, in the last row of Table 1, the temporal process is an operational process, and integration occurs as an accumulation of perceived objects: 1 R → 2 R → 3 R.

Information integration is rooted in the acquired knowledge and knowledge structure of a system, i.e., common features and their associations. At levels below the operator, integration is guided by descriptive knowledge, such as recognizing objects in images (as shown in Table 1) and drawing objects (as shown in Table 2). However, at the top level, occurring in the operator, integration is driven by temporally sequenced processes to either extract operational common features from a sequence of information or manipulate information based on the activation of known operational knowledge. Table 3 lists three examples in which the adder relies on its acquired operational knowledge to integrate information through sequential manipulation, performing addition and subtraction. In the symbol string "4 + 3 =" from the first example in Table 3, "4" and "3" represent two subsets of numeric counts, "+" triggers an mental operation to combine the two subsets, and "=" calls for the mental operation to count the final combined set.

## The reportability of mental states

The mechanism of reportability has not received as much attention in theories of consciousness and cognitive neuroscience explorations compared to most other easy problems. However, some influential theories of consciousness provide insights related to their core claims rather than addressing the mechanism itself. For instance, the GWTs suggest that mental states become reportable when they enter a global cognitive workspace, making them accessible for verbalization and decision-making, and the HOT posits that conscious reportability requires a mental representation of one's own cognitive states. Additionally, neuroscience research indicates that reportable consciousness is linked to activity in the PFC, parietal cortex, and thalamus (Koch and Tsuchiya 2007).

This adder is able to generate its semantic report and image report, as shown in Tables 1, 2, and 3. Its reportability is one of the fundamental capabilities of an executable system and is determined by two factors. First, the reported stimulus must be comprehensible to the system or related to its existing knowledge so that the associated symbol can be activated in the semantic report, or an object with common associated features can be activated in the image report. In other words, existing knowledge associated with the stimulus is a prerequisite for reportability. Similar to information integration and the ability to discriminate and categorize, reportability is an intrinsic property of its semantic knowledge. The second factor is related to information processing stages. As shown in Fig. 1, this adder can generate its report only after the associated information has been processed by the operator and projected back to the SS (to activate symbol output) and FS (to activate image output). This specific processing stage for a report to occur aligns with the empirical finding that reportable consciousness is associated with activations in the PFC and two other regions (Koch and Tsuchiya 2007). It appears to also align with some versions of GWT, in which the global workspace is considered to be related to the PFC.

**The ability of a system to access its own internal states**

This ability to access its own internal states pertains to the cognitive process "by which information about internal states is retrieved and made available for verbal report," as specified by Chalmers. This ability is one of the focal points of consciousness studies, with different theories offering distinctive perspectives on how a system accesses its internal states. In GWTs, access to internal states occurs when information is broadcast to a central global workspace; in HOTs, internal states become accessible when the brain generates higher-order representations of those states; in RPT, access is achieved through recurrent neural processing, where information flows back and forth between lower- and higher-level brain areas; and in IIT, access to internal states is an intrinsic property of highly integrated systems, where information is not only stored but also meaningfully connected.

In the adder, as shown in Fig. 1, information becomes available for further processing—i.e., information manipulation and reporting—only after it enters the mental operator. This mechanism is compatible with most influential theories, particularly some versions of GWTs, in which the PFC plays an important role in accessibility (Baars 1988, Dehaene and Changeux 2011, Mashour et al. 2020). It is also in line with HOTs, which emphasize higher-order thoughts, since processes in the operator constitute the top-level process in the adder; with RPT, since processing in the mental operator initiates feedback; and with IIT, since top-level integration occurs only in the operator of the adder, as discussed earlier.

**The focus of attention**

Attention is another focal point of consciousness studies, and various theories have proposed distinctive perspectives emphasizing different functions of attention, such as selecting, integrating, or amplifying information to make it available for conscious experience. In GWTs, attention functions as a "spotlight" that gates which information is broadcast across the brain; in HOTs, attention focuses on what higher-order thought processes deem most relevant; in RPT, attention results from recurrent processing that amplifies specific sensory inputs, making them dominant in conscious awareness; and in attention schema theory (AST) (Graziano 2017), attention is shaped by a self-model in which the brain uses attention schemas to manage and prioritize sensory input and cognitive tasks. According to GWTs, attention selects and amplifies specific signals, allowing them to enter the workspace (Mashour et al. 2020). In both GWTs and RPT, attention provides a selective boost to sensory signals, enabling them to reach the PFC and parietal regions, thereby engaging conscious access (Lamme 2010, Seth and Bayne 2022). Most theories addressing attention imply the assumption that attention is a separate neural mechanism from consciousness, aligning with growing evidence that distinct neural mechanisms underlie attention and consciousness (Maier and Tsuchiya 2021).

The adder doesn't have a designated region for attention; however, it suggests two mechanisms with functions partially similar to those of attention proposed in the theories reviewed above. One is the interface (elaborated as the borderline) of the mental operator, which filters incoming information from within. The borderlines of every functional region, as shown in Fig. 1, form its interface, depleted of information transformation and redistribution between regions. The other mechanism is the bias applied during information manipulation by the "emotion" region, which has also been previously discussed. These two qualitative perspectives highlight the importance of attention in information processing. In a biological brain, countless threads of information emerge and submerge at any given moment, requiring selection for high-level processing. Without the gating or filtering, a great amount of information overlays, overlaps, and joins into white noise. The PFC would be overwhelmed by the white noise and unable to function.

**The deliberate control of behavior**

There have been a number of competing theories addressing deliberate control. It is worth noting that, in most of these theories, deliberate control is dependent on the PFC, while the differences among them lie in the degree of this dependence and how the PFC interacts with other cognitive regions, such as the amygdala and basal ganglia. The prominent theories include: Executive control theories, developed as part of the theory of working memory, in which control is executed through the PFC (Baddeley 1996, Miller and Cohen 2001); dual-process theories, which distinguish between deliberate and automatic control (Kahneman 2011), self-regulation theories, which describe control through a feedback loop (Carver and Scheier 1998). Compared to control theories developed in other fields of research, this "easy problem" has not been substantially addressed in consciousness studies, including within these influential theories. It seems that most proposed models of consciousness do not explicitly discuss the deliberate control of behavior.

The adder's deliberate control is executed within the mental operator, aligning with all the cited theories in which the PFC plays the most important role in deliberate control. However, in the adder, deliberate control is not an independent mechanism but rather a part of its information integration process, as discussed in relation to the second easy problem, similar to accessibility to internal states. Since any process occurring within the mental operator is part of information integration, regardless of its consequences for learning or information manipulation, deliberate control is intended to influence the outcome of these processes within the operator and is therefore part of information integration.

Two kinds of deliberate control are proposed to occur in the operator, as indicated in Fig. 1. The first is an implemented mechanism that utilizes its acquired operational knowledge of semantic memory. Operational knowledge consists of rules or common features of information manipulation acquired through experience, which are called upon during intentional cognitive processes (e.g., addition or subtraction), as demonstrated in Tables 1, 2, and 3. The second kind involves biases applied to both the SS and FS buffers by mental states arising from the emotion region. However, deliberate control by emotion is not an implemented function but rather a perspective that highlights the importance of emotion in biasing problem-solving and decision-making in organisms.

**The difference between wakefulness and sleep**

Although this easy problem has also not been on the radar of most consciousness theories, the differences between wakefulness and sleep have been extensively studied, particularly in dream sleep—a unique subjective conscious experience during sleep. The sleep state is divided into two phases: REM (rapid eye movement) sleep and NREM (non-REM) sleep. NREM and REM sleep alternate throughout the entire sleep state. Dreams are reported in 70–95% of awakenings from REM sleep, compared with 5–10% from NREM, whereas mentation recall occurs in 43–50% of NREM awakenings (Foulkes 1962, Hobson 1988, Nielsen 2000). In dream sleep, sensory inputs are cut off, the neocortex is almost as active as in wakefulness, the limbic system is intensely active, and the PFC is less active (Maquet et al. 1996, Braun et al. 1997). Dreaming generally lacks self-reflection and self-control, and the dreamer rarely considers the possibility of actually controlling the flow of dream events (Purcell et al. 1986). During dream sleep, recent waking patterns stored in the hippocampus are reactivated in segments (Pavlides and Winson 1989, Fosse et al. 2003). Ample evidence from neurophysiological and psychological studies on dream states and dream reports has led to the well-received activation-synthesis model (Hobson 1988, Hobson et al. 2000), which proposes that dreams result from random impulses and are the consequence of the forebrain trying its best to make sense of these random impulses.

Regardless of the seemingly drastic difference between waking and dream sleep, a great number of empirical studies have concluded that dream sleep may play an important role in learning, problem-solving, and memory consolidation (e.g., Pearlman 1971, Winson 1985, Hennevin et al. 1995).The question is how the apparent cognitive and psychological benefits of dream sleep arise from its randomness and lack of deliberate control. This raises the issue of whether randomly activated segments under minimal deliberate control can still lead to learning, problem-solving, and memory consolidation—a possibility that has been demonstrated in the cognitive system from which the adder is expanded (Zhang 2009a, 2009b). In short, although there are great differences between waking and dream sleep states, they have similar consequences for one's cognitive growth, and these consequences in both waking and sleep are complementary.

## Summary

This study, based on an executable cognitive system and its implemented computational mechanisms, provides a systematic interpretation of most of the easy problems that can be "explained in terms of computational or neural mechanisms" addressed by Chalmers. It can be concluded from the systematic answers that the mechanisms and interpretations offered in this study are primarily rooted in the semantic knowledge of the system and how the system utilizes this knowledge to interact with the external world. Since semantic knowledge comprises both descriptive and operational knowledge, the easy problems can be seen as inquiries into different characteristics of the semantic knowledge of cognitive systems. Thus, the mechanisms that address the easy problems represent different perspectives on the semantic knowledge of the cognitive system.

Being able to discriminate and categorize, and therefore react to stimuli, are capacities of the semantic knowledge of a cognitive system. They all share the same mechanism, which is intrinsic to the knowledge structure gradually formed through learning from experience. This mechanism is essential for acquiring new knowledge and applying known knowledge. Information integration is another inherent property of the semantic knowledge of a cognitive system, also gradually formed through learning from experience. It is essential for the abilities to discriminate and categorize. Reportability, access to internal states, deliberate control, and attention are all associated with the mental operator, enabling it to extract and apply operational knowledge to interact with the external world. Specifically, attention regulates what internal information is accessed; deliberate control manages how information is manipulated; the manipulated result exits the operator before is reported. Sleep, especially dream sleep, is a special state in which the cognitive system processes stimuli that are randomly activated segments of past experience under minimal deliberate control.